\documentclass[twoside]{article}
\usepackage{Proc_NTSE}
\usepackage{latexsym}
\usepackage{epsfig}
\usepackage{graphicx}
\usepackage[english]{babel}
\usepackage{epstopdf}
\usepackage{color}
\begin{document}
\thispagestyle{plain}

\begin{center}
{\Large \bf \strut
 Modeling nuclear parton distribution functions
\strut}\\
\vspace{10mm}
{\large \bf 
H.~Honkanen$^{a}$, M.~Strikman$^{a}$ and V.~Guzey{$^b$}}
\end{center}

\noindent{
\small $^a$\it The Pennsylvania State University, 104 Davey Lab, University Park, PA 16802, USA } \\
 {\small $^b$\it National Research Center ''Kurchatov Institute'', Petersburg
Nuclear Physics Institute (PNPI),
Gatchina, 188300, Russia}

\markboth{
H.~Honkanen, M.~Strikman and V.~Guzey}
{
Modeling nPDFS} 

\begin{abstract}

The presence of nuclear medium and collective  phenomena which involve  several
nucleons  modify the parton distribution functions of nuclei (nPDFs) 
compared  to those of a free nucleon. These modifications have
been investigated by different groups  using 
global analyses of high energy nuclear reaction world data resulting 
in modern nPDF parametrizations with error estimates, such as EPS09(s), 
HKN07 and nDS. These phenomenological nPDF sets
roughly agree within their uncertainty bands, but have antiquarks for
large-$x$ and gluons for the whole $x$-range poorly constrained by the 
available data. In the kinematics accessible at the LHC this has negative 
impact on the interpretation of the heavy-ion collision data, especially for 
the $p + A$ benchmarking runs. The
EMC region is also sensitive to the proper definition of $x$, where the nuclear
binding effects have to be taken into account, and for heavy nuclei 
one also needs  to take into account that  a fraction of the nucleus
momentum is carried by the equivalent photons which
 modifies the momentum sum rule. We study how these effects affect the 
predictions for the nuclear modification ratios at the LHC kinematics using 
a model where we combine theoretical input for the leading twist
nuclear shadowing (the FGS model)
and the EKS98s/EPS09s nPDF set where the spatial
dependence is formulated as a power series of the nuclear thickness functions 
$T_A$.
\\[\baselineskip] 
\end{abstract}

\section {Proper definition of $x$}

Nuclear parton distribution functions (nPDFs) are usually defined for each 
parton flavor in terms of  nuclear modifications
$R_i^A(x,Q^2)$ and the corresponding free proton PDF $f_i^p(x,Q^2)$ such that
\begin{equation}
f_i^A(x,Q^2)\equiv R_i^A(x,Q^2)\,f_i^p(x,Q^2) \,,
\end{equation}
where the Bjorken $x =A\,Q^2/(2 q\cdot p_A)$, with $0\le x\le A$. 
In the collider frame, $x$ is simply the fraction of the nucleus momentum 
scaled by the  factor $A$.
However, since these nPDF sets are built ``on top'' of proton PDF sets, the tail
$1\le x\le A$ is usually ignored.

Phenomenological parametrizations for nuclear modifications, such as EKS98 
\cite{Eskola:1998df}, EPS09 \cite{Eskola:2009uj},  HKN07 \cite{Hirai:2007sx}, 
DSZS \cite{deFlorian:2003qf}, nCTEQ \cite{Kovarik:2010uv}  etc.~are largely
based on deep inelastic scattering 
(DIS) data, which is given  as a function of
$x_p=Q^2/(2 q_0 m_p)$, which is independent of the target mass.
 The difference between $x_p$ and $x$  thus
originates from the nuclear binding (see Refs.
\cite{Frankfurt:2011cs,Frankfurt:2012qs}\footnote{In 
Ref~\cite{Frankfurt:2012qs} there is a
sign error in the corresponding formula, Eq.(20). For detailed discussion, see the revised version of [7].}
\begin{equation}
x_p=  x\,(1\,+r^A_x) \,,
\label{xp}
\end{equation}
where 
\begin{equation}
r^A_x = \frac{1}{m_p}\left((m_n-m_p)N/A-\epsilon_A\right)<0 \,.
\label{rx}
\end{equation}
As an example, the nuclear binding energy $\epsilon_A\approx 7.88\, (7.68)$ MeV for Pb (C). 

In addition to the nuclear binding energy, the 
fraction of nucleus momentum carried by equivalent photons has to be taken 
into account in high energy heavy nuclear collisions.
The fraction of the nucleus
momentum carried by the photons is found to be \cite{Frankfurt:2012qs}
 \begin{equation}
\eta_{\gamma}({}^{12}{\rm C})=0.11\%,\,\,\eta_{\gamma}({}^{208}{\rm Pb})=0.7\%.
\end{equation}
Since the gluon nPDFs  are least constrained 
by the DIS and DY data, presence of the photons in the momentum sum rule 
mostly affects the overall momentum carried by gluons. 
The effect of  the  equivalent photon  field can 
be taken into account by rescaling the gluons after Eq.(\ref{xp}) has been 
applied to satisfy 
\begin{equation}
\sum_i\int_0^1 \mathrm{d}x x f^A_{i}(x,Q^2)=1-\eta_{\gamma}(A).
\label{rescale}
\end{equation}
To apply the ``conventional'' nPDFs
given as a function of $x_p$
for the calculation of the nuclear effects in the  ultra-relativistic heavy ion collisions where 
$x$ is used, one has to  translate to the  ``conventional'' nPDFs, 
given as a function of $x_p$, by taking into account the difference between 
$x_p$ and $x$ 
which from now on we explicitly call $x_{\rm shift}$, as follows
\begin{equation}
x_{\rm shift}f_i^A(x_{\rm shift},Q^2)=\begin{cases} \frac{x_p}{1+r^A_x} f_i^A(\frac{x_p}{1+r^A_x},Q^2),\,\, i=q,\bar q \\
g_{\rm scale}\frac{x_p}{1+r^A_x} f_i^A(\frac{x_p}{1+r^A_x},Q^2), \,\, i=g \end{cases},
\label{xshift}
\end{equation}
where the scaling factor for gluons, $g_{\rm scale}$, is determined via
Eq.(\ref{rescale}). Note that for a free proton,
 $x=x_p=x_{\rm shift}$.
It follows from Eq.~(6) that while the fraction of the nucleus momentum
carried
by quarks
is invariant with
respect to
the $x_p \to x_{\rm shift}$ conversion, the amount of the nucleus
momentum carried
by gluons, $\eta_g \equiv \int^{1}_{0} dx_{\rm shift} x_{\rm shift}
g_A(x_{\rm shift},Q^2)$,
decreases by the factor of $g_{\rm
scale}=\frac{\eta_g-\eta_{\gamma}(A)}{\eta_g}$ with the rescaling.

\section {Theoretically motivated nPDF model}

In this work we combine a small-$x$ theoretical model for the leading twist
nuclear shadowing, the FGS model \cite{Frankfurt:2011cs}, with the 
phenomenological EKS98/EPS09 nPDF set. The FGS model is based on 
the generalization of the Gribov-Glauber multiple scattering formalism and 
QCD factorization theorems. Using the picture of high energy scattering in the 
laboratory frame \footnote{The equivalent picture can be formulated in the 
nucleus fast frame \cite{Frankfurt:2011cs}.} and the notion of cross section fluctuations of 
energetic 
projectiles, multiple interactions are modeled using the effective 
$x$-dependent and flavor-dependent rescattering cross section 
$\sigma^i_{\rm soft}(x,Q^2)$, 
which controls the strength of the resulting nuclear shadowing.
In \cite{Frankfurt:2011cs}, based on the phenomenological analysis of cross 
section fluctuations in virtual photons, two models were 
suggested: model 1 (here  referred to as FGS1) and
model 2 (FGS2) corresponding to the upper and lower bounds on the 
predicted nuclear shadowing, respectively.
Both of the models were built on top
of CTEQ5 PDFs \cite{Lai:1999wy} (given as a function of $x_p$), and we will 
use this set for the combined
model as well. In this paper,
we will work in LO.

In the following the initial scale sea quark and gluon ($Q_0^2=2.5$ GeV$^2$) 
nuclear modifications for the region $10^{-4}\le x\le 0.01$,
where data practically do not constrain nPDFs, are taken from the FGS1 and FGS2
parametrizations;  for $0.03\le x\le 1.0$, the nuclear modifications are
taken from the EKS98 parametrization \cite{Eskola:1998df} (which in this 
region is very similar to the newer set EPS09 \cite{Eskola:2009uj}). For the
valence quarks, the  nuclear modifications are taken from EKS98 for the 
whole $x$-range.
The two parametrizations are combined by performing polynomial
interpolation between them, and (after being corrected for the
difference in the argument according to Eq.~(\ref{xp})) the gluons are
rescaled as in Eq.~(\ref{xshift}).
The resulting scaling factor at $Q_0^2=2.5$ GeV$^2$
for the gluons is $g_{\rm scale}\sim 0.984$ 
($g_{\rm scale}\sim 0.997$) for Pb (C) nucleus.  
(The 
change
 in the scaling factor is $~0.4\%$ at
$Q^2=10000^2$ GeV$^2$, so in practice a uniform scaling factor can be used
for any scale.)
Consequently the amount of the
momentum carried by the gluons decreases by 0.88\% (0.14\%)
for Pb (C). The fraction of the momentum carried by the gluons for each model 
is listed in Table~1 (41.80\% for the proton in CTEQ5L).
\begin{table}[h]
\label{momtab1e}
\caption{The percentage of nucleus momentum carried by the gluons.}
    \begin{center}
\begin{tabular}{cccc}
\hline\noalign{\smallskip}
& EKS98 & FGS1+EKS98  & FGS2+EKS98 \\
\noalign{\smallskip}\hline\noalign{\smallskip}
&{Pb \hspace{0.5cm} C} & {Pb \hspace{0.5cm} C} & {Pb \hspace{0.5cm} C}  \\
\noalign{\smallskip}\hline\hline\noalign{\smallskip}
$x_p$:  & {43.98 \hspace{0.2cm} 42.61} &  {44.20  \hspace{0.2cm}  42.58} & {44.30 \hspace{0.2cm} 42.62}   \\
\noalign{\smallskip}\hline\noalign{\smallskip}
$x_{\rm shift}$: & {43.58  \hspace{0.2cm} 42.54 }& {43.81 \hspace{0.2cm} 42.52}  & {43.91 \hspace{0.2cm} 42.56} \\
\noalign{\smallskip}\hline
\end{tabular}
\end{center}
\end{table}

\begin{figure}[h]
\begin{center}
\includegraphics[width=9.5cm]{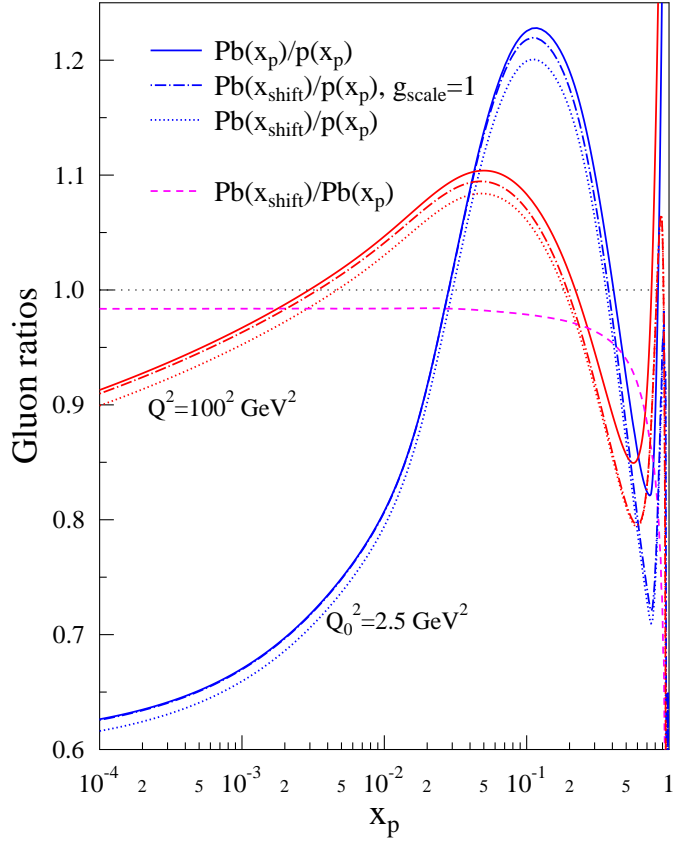}
\caption[a]{\protect \small Gluon ratio for 
${\rm Pb}(x_{\rm shift})/{\rm Pb}(x_p)$ at $Q_0^2=2.5$ GeV$^2$ and for 
${\rm Pb}/p$ at $Q_0^2=2.5$ GeV$^2$ and $Q^2=100^2$ GeV$^2$ for EKS98. }
\label{gluon_ratio_p_play.eps}
\end{center}
\end{figure}

In Fig.~\ref{gluon_ratio_p_play.eps} we show how changing the definition of $x$ 
affects the EKS98 gluon 
modification ratio ${\rm Pb}/p$ at the initial scale $Q_0^2=2.5$ 
GeV$^2$ and how this 
effect evolves up to the higher scale $Q^2=100^2$ GeV$^2$. In this work 
we use the QCDNUM DGLAP evolution code \cite{Botje:2010ay}. 
 At the initial scale
the original EKS98 gluon modification ratio 
$x_pG^{\rm Pb}(x_p)/x_pG^p(x_p)$ (solid line) is first modified 
setting $g_{\rm scale}=1$ in Eq.~(\ref{xshift}) (dotted-dashed line).
 As a result, the gluon modification ratio is only essentially modified at the EMC-region, 
$x_p>0.5$, where the (n)PDFs  are decreasing rapidly. For the full conversion 
with $g_{\rm scale}\sim 0.98$  (dotted line), the gluon nPDF is naturally scaled 
down over the whole $x_p$-range.
When evolved up to $Q^2=100^2$ GeV$^2$, the differences persist and spread
towards smaller values of $x_p$.
Using $x_{\rm shift}$ instead of $x_p$ obviously affects gluons for the whole 
$x_p$-range due to the rescaling (and at  higher scales also the sea quarks via
the DGLAP evolution), but for all the parton flavors the most
prominent effect sets in at the EMC-region, where the parton distribution
functions change quickly. 
As seen from above, the gluon ratio ${{\rm Pb}(x_{\rm shift})}/{\rm Pb}(x_p)$ (dashed line) 
is $<1$ for the whole $x_p$-range.

Note in passing that even for the valence quark distributions, the 
experimental information on the EMC effect in the region where the leading 
twist contribution dominates is very limited as the higher twist effects give 
a large (dominant ?) contribution to the $eA$ scattering cross section for 
$x\ge 0.5$ in the SLAC and JLab kinematics. Hence,
to date, practically no data on the quark modification for $x\ge 0.5$ in the scaling 
region is available for heavy nuclei such as, e.g., lead.

\begin{figure}[h]
\begin{center}
\includegraphics[width=9.5cm]{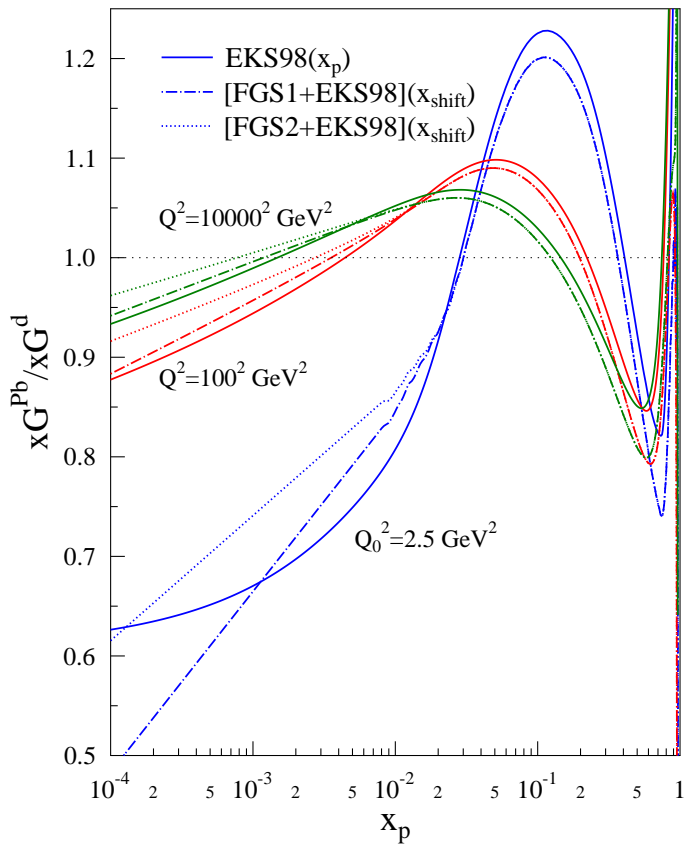}
\caption[a]{\protect \small Gluon ratios for  ${\rm Pb}/d$ 
at $Q_0^2=2.5$, 100$^2$ and 10000$^2$ GeV$^2$  for EKS98, and
the combination model FGS1(2)+EKS98.}
\label{gluon_ratio_d.eps}
\end{center}
\end{figure}

In Fig.~\ref{gluon_ratio_d.eps} we 
show the gluon ratio (${\rm Pb}/d$)
for the combined models FGS1+EKS98 (dotted-dashed) and FGS2+EKS98 (dotted), 
together with the EKS98 (solid) gluon modification. At the initial scale 
above $x_p>0.01$, the difference between the models originates 
only from the different definition of $x$; below
$x_p\le 0.01$, the two FGS nuclear shadowing models span a region considerably 
smaller than the error band of EPS09 gluons (see 
\cite{Eskola:2009uj} for details). 
When evolved up to $Q^2=10000^2$ GeV$^2$, a relevant scale in the LHC 
kinematics, it is evident that processes which are sensitive to the EMC-region 
are also sensitive to the proper definition of $x$.

\section{Consequences for the LHC}

In order to understand the sensitivity of the LHC kinematics to the EMC effect,
we study inclusive $\pi^0$ production in $p+{\rm Pb}$ collisions, which 
schematically can be expressed as  
\begin{eqnarray}
&& \sigma^{p+{\rm Pb} \to \pi^0 + X} = \nonumber \\
&& \sum_{i,j,k=q,\bar{q},g} {f_i^p}(x_1,Q^2) \otimes {f_j^{\rm Pb}}(x_2,Q^2) \otimes \hat{\sigma}^{ij\to k + X}(x_1,x_2,Q^2) \otimes D_{k \to \pi^0}(z,\mu_F^2),
\end{eqnarray}
where the factorization and renormalization scales have been set 
equal (see e.g.~\cite{Eskola:2002kv} for the formulae and details). 
In this work, we choose $\mu_F=p_T$ (the outgoing pion transverse momentum) 
and $Q=q_T$ (partonic transverse momentum). 

\begin{figure}[h]
\begin{center}
\includegraphics[width=12cm]{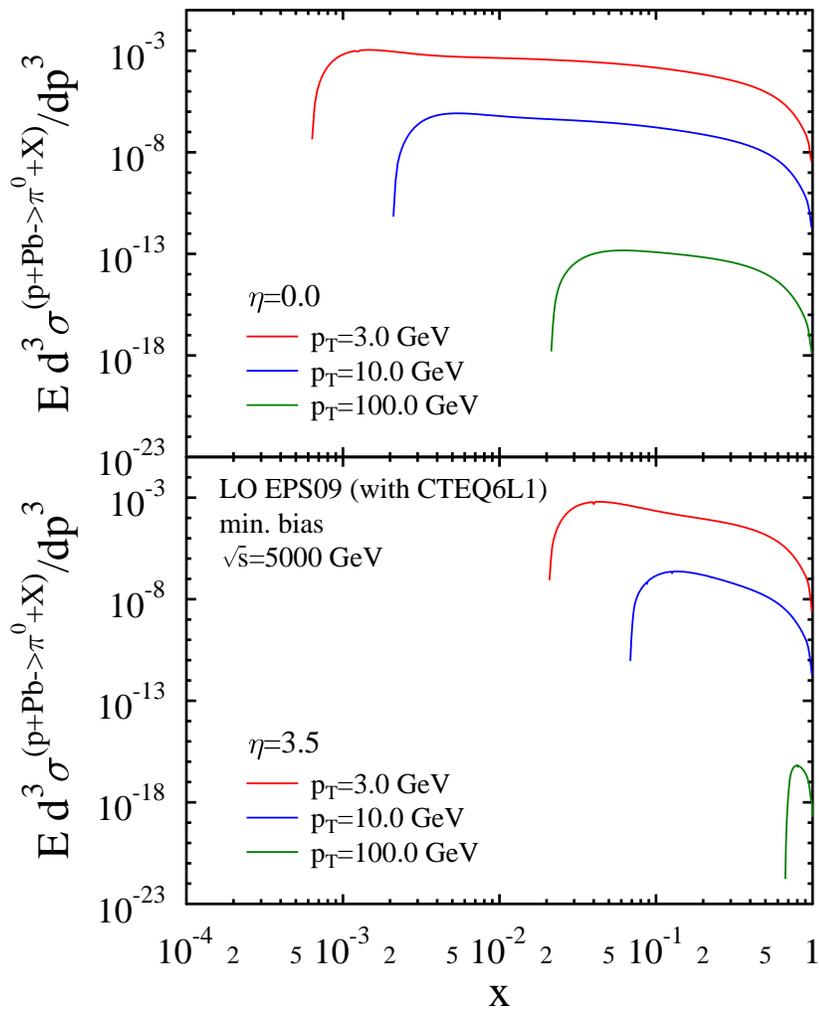}
\caption[a]{\protect \small The $x$-distribution for the minimum bias $\pi^0$
 production at the LHC at $\eta=0$ and $\eta=3.5$ for different $p_T$.} 
\label{Pb_eta_0.0_xbin.eps}
\end{center}
\end{figure}

In Fig.~\ref{Pb_eta_0.0_xbin.eps} we
show the LO invariant cross section $E d^3\sigma/dp^3$ for 
${p+{\rm Pb}\to\pi^0+X}$ at $p_T=3.0,\, 10.0$ and 100.0 GeV as a function
of $x_2$ (the momentum fraction carried by the parton in Pb, without conversion
given in Eq.~(\ref{xshift})). 
The results have been computed with the EPS09 
nuclear modifications \cite{Eskola:2009uj}, CTEQ6L PDFs \cite{Pumplin:2002vw} 
and DSS fragmentation functions \cite{deFlorian:2007aj}. 
Working in LO the overall normalization of the spectra is not fixed, 
but the $x_2$-distribution and the relative normalization  are not affected 
by this.
The upper panel shows the mid-rapidity and the lower panel -- the forward  
rapidity (at the LHC, the Pb rapidity is positive) $\pi^0$ production. For each 
$p_T$ value studied, the mid-rapidity 
production peaks at an order of magnitude smaller values of $x_2$ than the
forward rapidity results, and  remains
significant over a wider range of $x_2$.
In the forward direction the  pion production is concentrated on a rather
narrow $x_2$-range, making it a more sensitive probe of nuclear effects. In 
particular, at high-$p_T$, the pions are produced exclusively from the 
EMC-region,
making them also sensitive to the definition of $x_2$. 

\begin{figure}[h]
\begin{center}
\includegraphics[width=12cm]{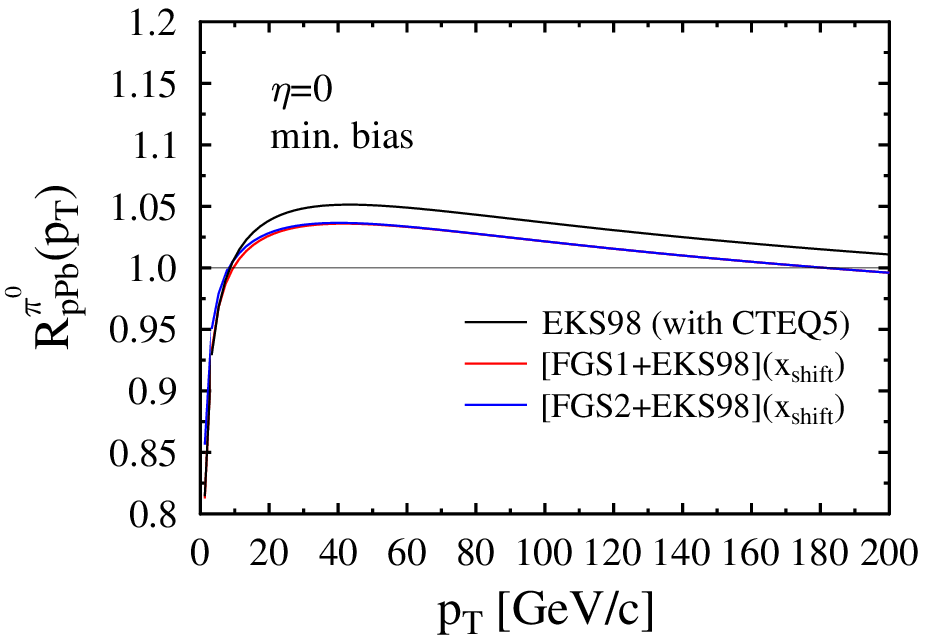}\vspace{-0.2cm} 
\includegraphics[width=12cm]{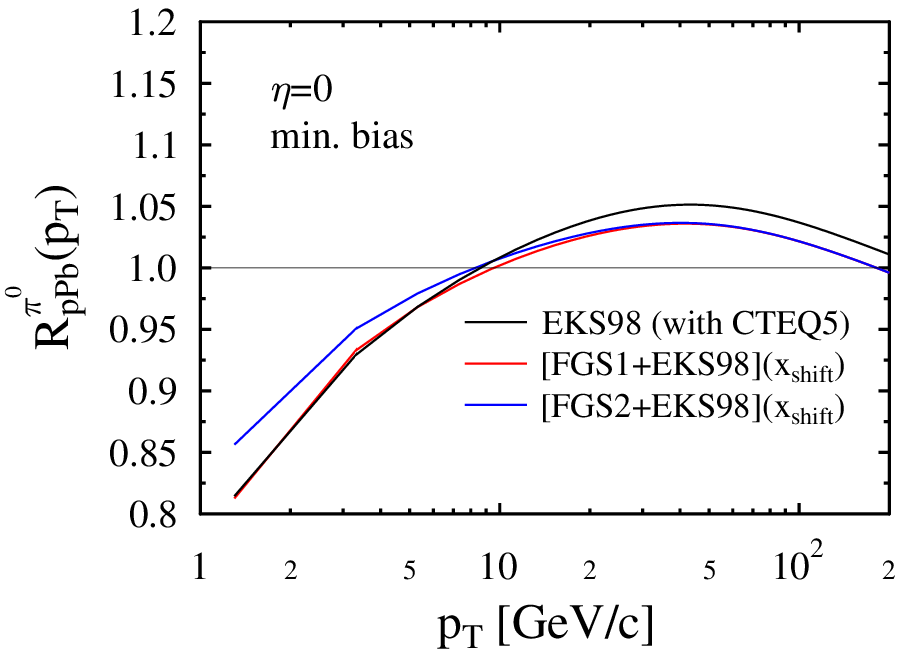}\vspace{-0.2cm}
\caption[a]{\protect \small Minimum bias FGS1(2)+EKS98 results for $R^{\pi^0}_{p{\rm Pb}}(p_T)$  for the LHC at $\sqrt{s}=5500$ GeV and $\eta=0.0$.
For comparison, the EKS98 grid result with CTEQ5L pdfs is also shown.
Upper panel on linear scale, lower panel on logarithmic scale.} 
\label{pPb_per_pp_mb_res.eps}
\end{center}
\end{figure}

\begin{figure}[h]
\begin{center}
\includegraphics[width=12cm]{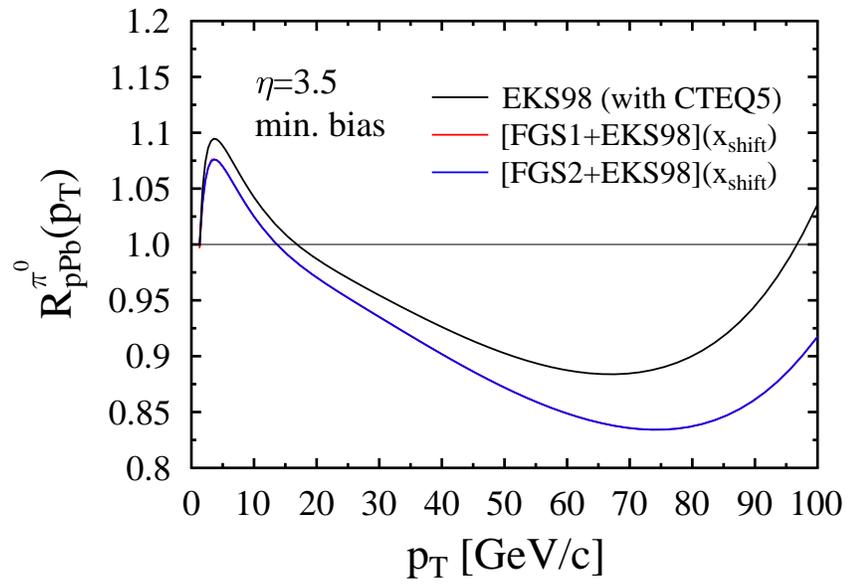}
\caption[a]{\protect \small Minimum bias FGS1(2)+EKS98 results for$R^{\pi^0}_{p{\rm Pb}}$  for the LHC at $\sqrt{s}=5500$ GeV and $\eta=3.5$.
For comparison, the EKS98 grid result with CTEQ5L pdfs is also shown. The FGS1+EKS98 and FGS2+EKS98 curves are 
indistinguishable.} 
\label{pPb_per_pp_mb_res_eta_3.5.eps}
\end{center}
\end{figure}

Figures~\ref{pPb_per_pp_mb_res.eps} and \ref{pPb_per_pp_mb_res_eta_3.5.eps} 
show  the FGS1(2)+EKS98 results for the minimum bias nuclear modification ratio,
\begin{equation}
R^{\pi^0}_{p{\rm Pb}}(p_T,\eta)=\frac{d^3\sigma^{p{\rm Pb}}/dp^3}{d^3\sigma^{pp}/dp^3} \,,
\end{equation}
 at the LHC at $\sqrt{s}=5500$ GeV. For comparison, the EKS98 results with 
the CTEQ5L PDFs are also shown. In the mid-rapidity 
(Fig.~\ref{pPb_per_pp_mb_res.eps}) the difference between the EKS98 and 
FGS+EKS98 results remain within a few-\%. As can be 
seen from  Fig.~\ref{Pb_eta_0.0_xbin.eps} (upper panel), up to $p_T\sim 100$ 
GeV, the pion spectra mostly  originate
from the gluon distribution dominated $x$-range
below the EMC-region. Therefore the differences between the models are caused
both
by the different assumptions about shadowing (see the lower panel where the
small-$p_T$ part is shown on a logarithmic scale) and the scaling of the gluon 
distribution with $g_{\rm scale}<1$.
of $x$. The difference between the models remain moderate even above 
$p_T>100$ GeV.

As seen from Fig.~\ref{pPb_per_pp_mb_res_eta_3.5.eps},
the situation is drastically different in the forward  rapidity pion 
production. Above $p_T\sim 10$ GeV, the two FGS-models are 
indistinguishable, but start to deviate  from the EKS98 result as 
$p_T$ increases. This sizable effect is caused by the correction to the 
 $x$-definition alone.

\begin{figure}[h]
\begin{center}
\includegraphics[width=16cm]{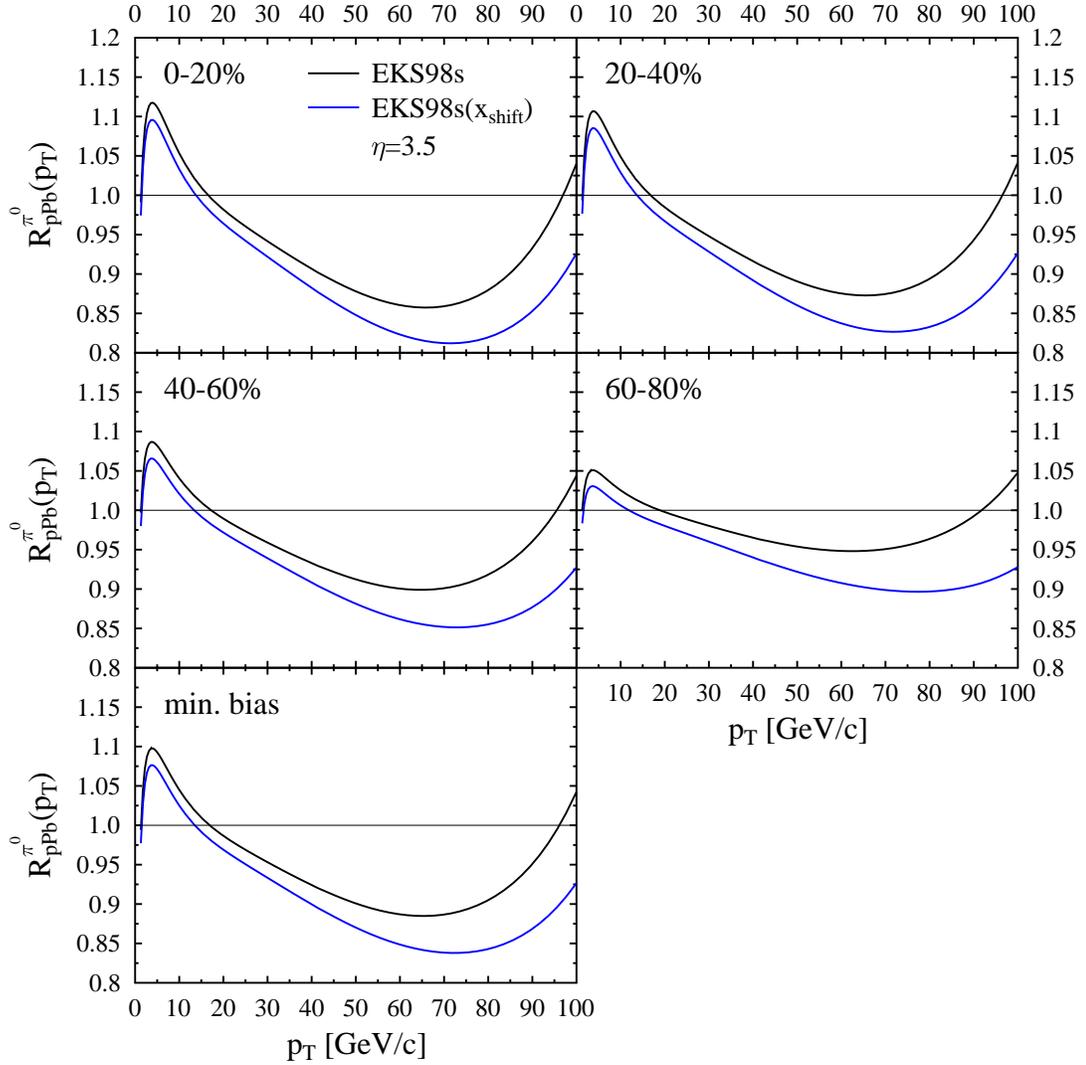} 
\caption[a]{\protect \small $R^{\pi^0}_{p{\rm Pb}}$  for the LHC at $\sqrt{s}=5500$ GeV and $\eta=3.5$ for the EKS98 (with CTEQ6L1) results 
computed with $x_p$ and $x_{\rm shift}$. The different centrality classed are 
computed as in
\cite{Helenius:2012wd}. The FGS1+EKS98 and FGS2+EKS98 curves are indistinguishable.} 
\label{pPb_per_pp.eps}
\end{center}
\end{figure}

Until now we have discussed the minimum bias results, where the impact
parameter dependence of the nuclear effects has been spatially averaged.
In the FGS model the transverse position ${\bf s}$ dependence is naturally 
built in  as functions of $T_A({\bf s})$ since the nuclear shadowing is first 
calculated for fixed ${\bf s}$  and next the integral over ${\bf s}$ is taken. 
In the EKS98s/EPS09s model \cite{Helenius:2012wd}, the EKS98 and EPS09 nPDF 
parametrizations were
also assumed to have spatial dependence as a power series of $T_A({\bf s})$. 
With the centrality classes modeled using the optical Glauber model,
the EPS09s results were  found to
be consistent with the mid-rapidity 
PHENIX $R_{d{\rm Au}}^{\pi^0}$ centrality systematics \cite{Adler:2006wg}.
However, as already seen in Fig.~\ref{pPb_per_pp_mb_res_eta_3.5.eps}, 
the proper definition of $x$ has a major effect on the LHC predictions
at forward rapidity. In Fig.~\ref{pPb_per_pp.eps} we applied
the procedure described in \cite{Helenius:2012wd} to 
$R_{p{\rm Pb}}^{\pi^0}$ at $\eta=3.5$ for a selection of different centrality
classes, and compared the EKS98s results (with CTEQ6L PDFs) with and
without the $x$-corrections. Irrespective of the centrality, the correction
causes a clear, measurable effect.

For an impact parameter dependent theoretically motivated nPDF
model it is also important to pay special attention to the EMC region for 
another reason. In \cite{Weinstein:2010rt} the magnitude of the EMC effect
was shown to be linearly related to the short range correlations (SRC) scale
factor measured from electron inclusive scattering at $x\ge 1$.
Consequently the impact parameter dependence of the EMC effect should be 
proportional to the local density \cite{Frankfurt:1988nt} and the 
EMC effect thus should be  strongest in the center of the nucleus. 
We will address this issue in \cite{upcoming}, where
a full NLO impact parameter dependent nPDF set will be released.

This work was supported in part by US DOE Contract Number DE-FG02-93ER40771. 



%

%

\end{document}